\documentclass{emulateapj}

\usepackage{epstopdf}
 
\newcommand{\re}{R$_{\rm e}$}
\newcommand{\atl}{ATLAS$^{\rm 3D}$}
\newcommand{\msun}{\hbox{$M_\odot$}}

\begin{document}

\title{Initial Mass Function for Massive Galaxies at z$\sim$1}

\author{Shravan Shetty$^1$ \and Michele Cappellari}
\affil{Sub-department of Astrophysics, Department of Physics, University of Oxford, Denys Wilkinson Building, Keble Road, Oxford \\
 OX1 3RH.\\
$^1$Email: {\tt shravan.shetty@astro.ox.ac.uk}} 

\maketitle

\label{firstpage}

\section{Abstract}

We present the results on the stellar Initial Mass Function (IMF) normalisation of 68 massive ($M_\ast=10^{11}-10^{12}$\msun) Early-Type Galaxies (ETGs) at redshift of $\sim$1. This was achieved by deriving the stellar Mass-to-Light ratio (M/L) of the galaxies through axis-symmetric dynamical modelling and comparing it to the same derived via stellar population modelling through full spectrum fitting. The study also employs an Abundance Matching technique to account for the dark matter within the galaxies. The results demonstrate that massive ETGs at high redshifts on average have a Salpeter-like IMF normalisation, while providing observational evidence supporting previous predictions of low dark matter fraction in the inner regions (\textless 1\re) of galaxies at higher redshift.

\section{Introduction}

In recent years, a significant paradigm shift has taken place in our understanding of the Initial Mass Function (IMF) of galaxies. In \citet[Hereafter \atl XV]{atlas3d15}, the authors created detailed axis-symmetric dynamical models to robustly derive the stellar Mass-to-Light ratio (M/L) for 260 ETGs in the local universe. By comparing these M/Ls to those derived through stellar population modelling, the authors demonstrated that the IMF of galaxies wasn't universal, but in fact varied systematically with the central velocity dispersion of the galaxies \citep[Hereafter \atl XX]{cappellari2012nature,atlas3d20}. This systematic variation has been confirmed via stellar population models \citep{Spiniello2012,Conroy2012,Smith2012,ferremateu2013,LaBarbera2013} and dynamical approaches \citep{Dutton2013,tortora2013}. In this work, we have attempted to study the IMF normalisation of high redshift galaxies. 

\section{Data}

We use 1D galaxy spectra taken from the DEEP2 Galaxy Redshift Survey \citep{DEEP2}. The survey has observed $\sim$49,000 galaxies in an observed-frame wavelength range of 6,500-9,100\AA ~at a resolution of R$\sim$6,000 at 7,800\AA. We also use high-resolution (0.03''/pixel) F814W filter HST/ACS images of our galaxies, taken from the AEGIS data product \citep{aegis}, which translates to the rest-frame B-band for our galaxies.

\section{Method}

Before we start our analysis we apply our first set of selection criteria. We restrict our analysis to galaxies in the DEEP2 survey that have reliable redshift and lay in the redshift range of 0.7-0.9. Along with this, we also visually inspect the galaxy spectra to verify the presence of significant absorption lines, and to mask telluric features and significant gas emission lines.

We derive the stellar kinematics of the galaxies by fitting the galaxy spectra to a subset of 53 empirical stellar spectra taken from the Indo-US Library of Coud\'{e} Feed Stellar Spectra Library \citep{indo-uslibrary}, using the pPXF code \citep{ppxf}. We have set a Signal-to-Noise (S/N) cutoff at 3 per 60 km/s spectral pixel and have visually inspected every spectrum to ensure that the best-fits do indeed fit real stellar features. We have derived realistic errors using a bootstrapping technique. 

To create our dynamical models, we use the JAM dynamical modelling code which solves the anisotropic Jean's equations under the assumption of axis-symmetry \citep{jam}, while taking into account aperture and seeing effects. Using the light and mass profile for the galaxy, the code is able to predict the expected observed stellar kinematics for the galaxy profiles which can then be calibrated with the observed kinematics to derive the stellar M/L of the galaxies. 

To parametrise the light and stellar mass profile of our galaxies, we use the MGE fitting code \citep{cappellari2002} on our galaxy photometry. For our galaxies, it is essential that we accurately reproduce the light profile of the inner region of galaxies, where the stellar kinematics are derived, and hence we have visually verified each galaxy fit. In the process of visual verification, we have eliminated galaxies that have non-axis-symmetric features such as bars, dust lanes etc as these aren't well parametrised by the code. 

\begin{figure}
\centering
\includegraphics[width=0.95\columnwidth]{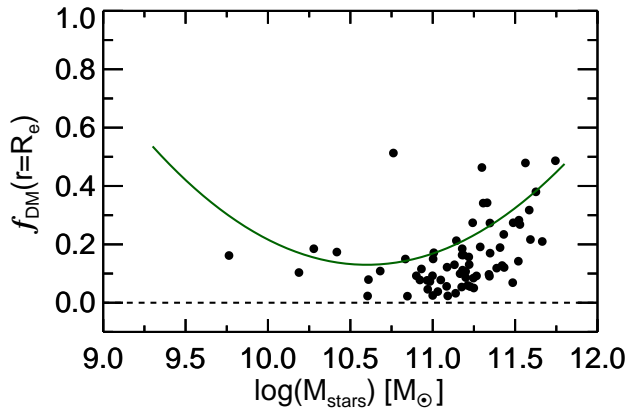} 
\caption{This plot depicts the dark matter fraction within 1\re ~of our galaxies and it's systemmatic variation with galaxy stellar mass. The solid line represents the trend found by \atl XV (Bottom panel of fig.~10). The plot above is directly comparable to that in \atl XV since both were produced using a similar Abundance matching technique.}
\label{Shetty_fig1}
\end{figure}

To model the halo profile, we have used results from Abundance Matching. To do this, we associate every possible stellar mass, or stellar M/L, of the galaxy with an associated halo mass at the given redshift of the galaxy using the results of \citet{MosterNaabWhite2013}. Assuming that the dark matter halo follows a NFW profile \citep{NFW_DarkHalo_1996}, we use the results of \citet{KlypinTrujillo-GomezPrimack2011} to derive the concentration of the profile, while accounting for the redshift of the galaxies, and thereafter other physical parameters of the halo. We can then derive the best fitting stellar mass, ie stellar M/L, which along with it's associated dark matter halo best reproduces the observed stellar kinematics of our galaxies through our dynamical modelling.

To use the JAM code we have to make certain assumptions on our galaxies, ie. we assume a constant velocity anisotropy ($\beta_{\rm z}\equiv1-\sigma_z^2/\sigma_R^2$) of 0.2 for our galaxies, and assume that the inclination of our galaxies is $60^\circ$ unless this is unrealistic given the galaxy photometry. These assumptions are supported by results of realistic simulations carried out by \citet{lablanche2012} and previous works \citep{cappellari2006,jam} that have shown that the errors introduced by the uncertainty in these quantities are insignificant compared to the errors on the stellar kinematics of our galaxies.

Next, we derive the stellar M/Ls through full spectrum fitting of the galaxy spectra with a template model set using pPXF. We use the MILES stellar population models \citep{milesmodels}, which are based high resolution empirical stellar spectra of the MILES stellar library \citep{mileslibrary}, under the assumption of a Salpeter IMF. We also assume that the star formation history of the galaxies varies linearly in log-time space. 

To implement our final selection criterion, we visually inspect the stellar populations of our galaxies and remove those that contain multiple major star formation events or significant fraction of young (\textless 1.2Gyr) stellar population. This criterion is based on results of \atl XV where the authors find that such young galaxies tend to have a radially varying stellar M/L, and hence break the implicit assumption of constant M/L throughout the galaxy. 

For further information on the methodology used and examples plots of the various fits in the work, we refer the reader to \citet{ShettyCappellari2014ApJL}, where the same analysis is done with the exception of Abundance Matching.

\section{Results and Conclusion}

The work presented here extends to higher redshift earlier works of the IMF normalization of local galaxies. Through a rigorous selection criteria, we have carefully derived the stellar M/L for 68 galaxies, from an original sample of $\sim$49,000 galaxies, through axis-symmetric dynamical modelling and stellar population modelling. As a result of these selection criteria, our galaxy sample is biased towards massive ($M_\ast=10^{11}-10^{12}$ \msun) ETGs.

The dynamical M/L derived in this work represents the stellar M/L of the galaxy as it takes into account the contribution of the dark matter halo through Abundance Matching. Based on the results of detailed dynamical modelling in \atl XV and on merger computer simulations of \citet{Hilz2013}, \citet{ShettyCappellari2014ApJL} had assumed that the dark matter fraction within the inner 1\re ~of these galaxies is \textless 30\%. The abundance matching technique used in this study supports this prediction. In Fig.~\ref{Shetty_fig1}, we depict the dark matter fraction of our galaxies. Here we notice a trend of the dark matter fraction with the stellar mass of the galaxies, which is similar to that observed by \atl XV (Bottom panel of fig.~10), with a lower normalisation, consistent with the results of \citet{Hilz2013}.

\begin{figure}
\vspace{10pt}
\centering
\includegraphics[width=0.95\columnwidth]{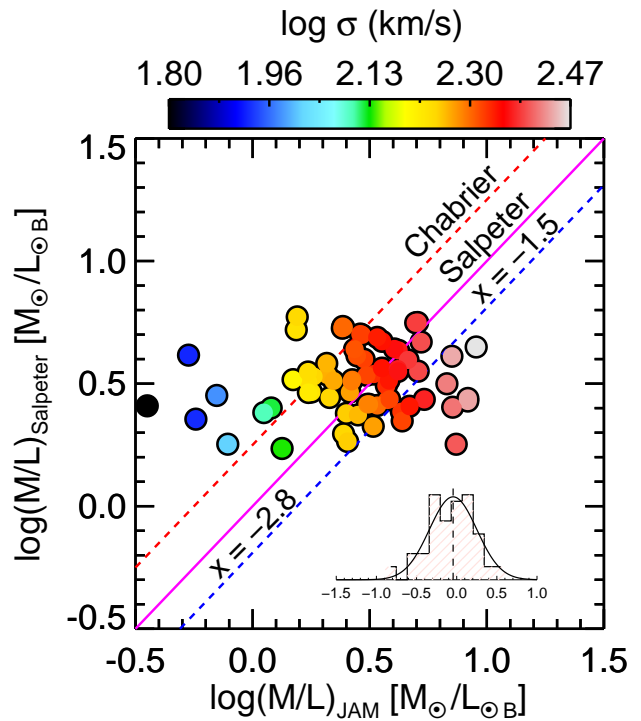} 
\caption{This plot illustrates the IMF normalisation of the galaxies. The x-axis represents the stellar M/L derived via our dynamical models, while the y-axis represents that derived through stellar population modelling. The plot is is directly comparable to fig.11 in \atl XX.}
\label{Shetty_fig2}
\end{figure}

In Fig.~\ref{Shetty_fig2}, we present the IMF normalisation of our galaxies. The colour code in the plot represents the smoothed variation of the central velocity dispersion of the galaxies. The plot clearly illustrates that, on average, the IMF normalisation of the massive high redshift ETGs is Salpeter-like. This plot also qualitatively similar to fig.~11 of \atl XX, ie both plot illustrate a lack of correlation between the two stellar M/Ls for the most massive galaxies. Assuming the passive evolution of the central regions of these galaxies, this result is consistent with results produced by \atl, \citet[SLACS]{auger2010b}, \citet{Conroy2013} and \citet{Spinielloetal2014}.

\bibliographystyle{apj}

\end{document}